\documentclass[fleqn,usenatbib]{mnras}

\usepackage{newtxtext,newtxmath}

\usepackage[T1]{fontenc}

\DeclareRobustCommand{\VAN}[3]{#2}
\let\VANthebibliography\thebibliography
\def\thebibliography{\DeclareRobustCommand{\VAN}[3]{##3}\VANthebibliography}

\usepackage{graphicx}
\usepackage{amsmath}
\usepackage{subcaption}
\usepackage{scalerel}
\usepackage[dvipsnames]{xcolor}

\usepackage{graphicx}
\usepackage{amsmath}

\newcommand{\Msun}{M$_\odot$}
\newcommand{\tdelay}{$t_{\mathrm{delay}}$}
\newcommand{\angstrom}{\mbox{\normalfont\AA}}

\title[Colour distribution of simulated satellites]{Colour and infall time distributions of satellite galaxies in simulated Milky-Way analogs}

\author[Yue Pan et al.]{Yue Pan,$^{1}$\thanks{E-mail: explorerpan@uchicago.edu}
    Christine M. Simpson,$^{1,2}$\thanks{Present address: ALCF, Argonne National Laboratory, Lemont, IL 60439, USA}
    Andrey Kravtsov,$^{1,2,3}$
    Facundo A. Gómez,$^{4,5}$
    Robert J. J. Grand,$^{6,7}$
    \newauthor\
    Federico Marinacci,$^{8}$
    Rüdiger Pakmor,$^{9}$
    Viraj Manwadkar$^{1}$ and
    Clarke J. Esmerian,$^{1}$
\\
\\
$^{1}$Department of Astronomy \& Astrophysics, The University of Chicago, Chicago, IL 60637, USA\\
$^{2}$Enrico Fermi Institute, The University of Chicago, Chicago, IL 60637 USA\\
$^{3}$Kavli Institute for Cosmological Physics, The University of Chicago, Chicago, IL 60637 USA\\
$^{4}$Instituto de Investigación Multidisciplinar en Ciencia y Tecnología, Universidad de La Serena, Raúl Bitrán 1305, La Serena, Chile\\
$^{5}$Departamento de Astronomía, Universidad de La Serena, Av. Juan Cisternas 1200 Norte, La Serena, Chile\\
$^{6}$Instituto de Astrofísica de Canarias, Calle Vía Láctea s/n, E-38205 La Laguna, Tenerife, Spain\\
$^{7}$Departamento de Astrofísica, Universidad de La Laguna, Av. del Astrofísico Francisco Sánchez s/n, E-38206, La Laguna, Tenerife, Spain\\
$^{8}$Department of Physics \& Astronomy "Augusto Righi", University of Bologna, via Gobetti 93/2, 40129 Bologna, Italy\\
$^{9}$Max-Planck-Institut für Astrophysik, Karl-Schwarzschild-Str. 1, 85748 Garching, Germany\\
}

\date{Accepted XXX. Received YYY; in original form ZZZ}

\pubyear{2022}

\begin{document}
\label{firstpage}
\pagerange{\pageref{firstpage}--\pageref{lastpage}}
\maketitle

\begin{abstract}
We use the Auriga simulations to probe different satellite quenching mechanisms operating at different mass scales ($10^5 M_\odot \lesssim M_\star \lesssim 10^{11} M_\odot$) in Milky Way-like hosts. Our goal is to understand the origin of the satellite colour distribution and star-forming properties in both observations and simulations. We find that the satellite populations in the Auriga simulations, which was originally designed to model Milky Way-like host galaxies, resemble the populations in the Exploration of Local VolumE Satellites (ELVES) Survey and the Satellites Around Galactic Analogs
(SAGA) survey in their luminosity function in the luminosity range $-12 \lesssim M_V \lesssim -15$ and resemble ELVES in their quenched fraction and colour--magnitude distribution in the luminosity range $-12 \lesssim M_g \lesssim -15$. We find that satellites transition from blue colours to red colours at the luminosity range $-15 \lesssim M_g \lesssim -12$ in both the simulations and observations and we show that this shift is driven by environmental effects in the simulations. We demonstrate also that the colour distribution in both simulations and observations can be decomposed into two statistically distinct populations based on their morphological type or star-forming status that are statistically distinct.  In the simulations, these two populations also have statistically distinct infall time distributions. The comparison presented here seems to indicate that the tension between the quenched fraction in SAGA and simulations is resolved by the improved target selection of ELVES, but there are still tensions in understanding the colours of faint galaxies, of which ELVES appears to have a significant population of faint blue satellites not recovered in Auriga.
\end{abstract}


\section{Introduction}

One of the most fundamental properties of galaxies is their bimodal colour distribution. Galaxies generally fall into two types: a red sequence characterized by a lack of star formation, primarily composed of massive, quenched elliptical galaxies, and a blue cloud characterized by ongoing star formation, mainly composed of star-forming disk galaxies \citep{Strateva2001,Baldry2004,Bell2004,  Menci2005}. Recent studies in the Local Volume (LV) have reproduced this colour bimodality in satellite galaxies \citep{Carlsten2022}. Understanding the underlying physical mechanisms that cause this colour bimodality in dwarf galaxies is fundamental to our understanding of galaxy formation and evolution, because star-formation (SF) activity correlates with color and, according to the hierarchical structure formation theory, all galaxies have once been dwarf galaxies \citep{White.Frenk1991}.

However, there is ongoing debate about quenching processes in dwarf galaxies and in particular of dwarf satellites orbiting larger galaxies \citep[see, e.g.,][for a recent overview]{Sales.etal.2022}. Studies in the Milky Way (MW) have found that except for a few massive objects \citep{Lewis2007, Fraternali2009, Makarov2012, Karachentsev2015}, all satellites of the MW within its virial radius are quenched. Extending to the Local Group (LG), nearly all dwarf galaxies with $M_\star < 10^8 M_{\odot}$ that are satellites within 300 kpc of the MW or M31 have quiescent SF and little-to-no cold gas, but nearly all isolated dwarf galaxies (i.e. in the field) are star-forming and gas-rich \citep{Mateo1998, Grcevich.Putnam2009, Geha2012, Spekkens2014, Wetzel2015b, Putman2021}. This field-satellite dichotomy has been a strong indicator of environmental quenching \citep{Lin.Faber1983, McConnachie2012, Slater.Bell2014, Weisz2015, Wetzel2015b}: satellites are quenched by external processes once they enter the gravitational bounds of their massive hosts. 

Moving beyond the LG, in the Local Volume (LV, $\leqslant 10$ Mpc), \citet{Carlsten2022} found that the majority of low-mass satellites ($M_\star \lesssim 10^7 M_{\odot}$) are quenched. However, the Satellites Around Galactic Analogs (SAGA) survey \citep{Geha2017, Mao2021} found that most of the satellites within the virial radii of MW-like hosts in the same satellite mass range are actively star-forming, in stark contrast to LG and LV satellites \citep{Karunakaran2021}. The question of why there is such a big difference between satellites and isolated field dwarfs becomes urgent. \citet{Font2022} argued that by considering the differences in host mass distributions and observation selection effects, the huge discrepancy between the quenched fractions of low mass satellite galaxies in LG and isolated MW-like systems is significantly reduced. 

Thanks to efforts in the theoretical front, we can now simulate MW and isolated environments for satellite galaxies at unprecedented resolution to understand the different physical mechanisms that drive galaxy quenching. Various processes have been proposed to stop star formation in satellite galaxies. Reionization is proposed to be driving quenching in low-mass dwarf galaxies \citep{Dekel.Silk1986,Thoul.Weinberg1996,Gnedin2000,Mayer2001, Brown2014,Weisz2014,Fillingham2016,Tollerud2018,Rodriguez.Wimberly2019}. Galaxy interaction can also quench some satellite galaxies \citep{Pearson2016, Pearson2018}. Internal processes such as stellar winds and supernova feedback can remove some part of the gas, but it is said to be insufficient for completely quenching the satellite \citep{Agertz2013,Emerick2016}. 

External environmental processes such as ram pressure stripping - a process by which the cold gas of the satellites gets stripped when passing through the circumgalactic medium (CGM) of the host - are said to be the dominant quenching mechanism for satellite masses $10^5 M_{\odot}\lesssim M_\star \lesssim 10^8 M_{\odot}$ \citep{Gunn.Gott1972,Murakami.Babul1999, Tonnesen.Bryan2009, Bahe.McCarthy2015, Fillingham2016, Kazantzidis2017, Simpson2018, Digby2019, Fillingham2019}.  Ram pressure is consistent with the rapid quenching timescale ($1\sim 2$ Gyr) of these satellite galaxies upon infall \citep{Fillingham2015}. Tidal stripping can also boost the efficiency of ram pressure stripping by diminishing the overall gravitational potential of the satellite galaxy \citep{Mayer2006}. Because more massive satellites are better able to retain their gas reservoirs \citep{Simpson2018,Garrison-Kimmel2019b} compared to lower mass satellites when interacting with hosts of the same mass, environmental quenching is less efficient for more massive satellite galaxies.

Starvation or strangulation - a scenario in which gas accretion onto the satellite galaxy is stopped after infall - can quench more massive satellite galaxies $(M_\star \sim 10^8 - 10^{11} M_{\odot}) $ \citep{van.den.Bosch2008, McGee2014, Wheeler2014, Fillingham2015, Phillips2015, Davies2016, Trussler2020} and the timescale of starvation is comparable to the gas depletion timescale \citep{Huang2012,Wetzel2013,Wheeler2014,   Fillingham2015}.  For environmental processes, \citet{Garrison-Kimmel2019b} used the FIRE simulations to identify differences in histories between “satellite versus central” galaxies and in different environments “LG versus individual MW versus isolated dwarf central”. They found that	around individual MW-mass hosts, central galaxies in the “near field” have more extended SFH than their satellite counterparts: the former more closely resemble isolated (true field) dwarfs, but this difference is muted in LG-like environments, suggesting that the paired halo nature of LG may regulate star formation in dwarf galaxies even beyond the virial radii of the MW and M31. Moreover, \citet{Hausammann2019} used a wind tunnel and a moving box technique to simulate both the ram pressure and tidal forces, and they found that while ram pressure is very efficient at stripping the hot and diffuse gas of the dwarf galaxies, it can remove their cold gas $(T < 10^3 K)$ only in very specific conditions. 

In this paper, we use the Auriga project \citep{Grand2017}, a suite of 30 cosmological magneto-hydrodynamical zoom simulations of galaxy formation in Milky Way-mass haloes, to probe further into the quenching mechanisms of satellite galaxies. We start by comparing the Auriga simulations with observations in terms of luminosity function, quenched fraction, and colour--magnitude diagram to establish how the results from simulations compare with observations. We then draw a connection between the colour and infall time distributions in the simulations by exploring timescales associated with satellite evolution as measured in the simulations. Finally, we identify different quenching mechanisms operating on different mass scales. 

The structure of this paper is as follows. In Section~\ref{sec:methods}, we describe the Auriga simulations and the methods we used to obtain satellite NUV-$g$ colours in the Auriga simulations. In Section~\ref{sec:comparison}, we compare the Auriga simulations to the ELVES sample in terms of luminosity function, quenched fraction, and color--magnitude diagram. In Section~\ref{sec:timescales}, we use the Auriga simulations to determine different timescales associated with quenching and explore the possible connection between quenching timescales and the simulated mass, colour, and magnitude of satellites. In Section~\ref{sec:discussions}, we discuss the comparison between the Auriga simulations and observations from SAGA and ELVES, and with other simulations with different implementations of underlying physics. Finally, we present our conclusions in Section~\ref{sec:summary}.

\section{Methods}
\label{sec:methods} 
In this section, we introduce the Auriga simulations and the methods we used to obtain NUV$-g$ colours for satellites in the Auriga simulations.
\subsection{The Auriga simulations}
\label{sec:Auriga_simulations} 

In this study, we use the Auriga simulations \citep{Grand2017} -- a suite of cosmological zoom-in simulations of $\approx 10^{12}$ \Msun\ haloes designed to simulate the formation of MW-sized galaxies.  The Auriga simulations were run with the $N$-body+magneto-hydrodynamics code \textsc{AREPO} \citep{Springel2010, Pakmor2016} and include many of the physical effects important in galaxy formation, including gravity, gas cooling, magnetic fields, star-formation, energetic feedback from stars and black holes, and metal enrichment \citep{Grand2017}.

We focus most of our study on the original 30 haloes of the Auriga suite that have a baryon mass resolution of $\sim 5 \times 10^4$ \Msun\ and a minimum physical gravitational softening length of 369 pc after $z=1$ (that scales with the scale factor prior to $z=1$).  We also consider resimulations of 6 haloes with 8 times better mass resolution that have a baryon mass resolution of $\sim 6.7 \times 10^3$ \Msun\ and a minimum softening length of 185 pc.  Following the conventions of the Auriga project, we call the lower resolution simulations ``Level 4'' simulations and the higher resolution simulations ``Level 3.''  It has been demonstated that the properties of galaxies we focus on in this study are well converged at the Level 4 resolution with the Auriga model for interstellar medium, star formation, and feedback  \citep{Grand2021}.

We make use of the Auriga halo catalogues created during the simulations that identify dark matter haloes with a friends-of-friends algorithm \citep{Davis1985} and gravitationally bound subhaloes identified with the  \textsc{SUBFIND} code \citep{Springel2001}.  We track inheritance between subhaloes with the merger tree code  \textsc{LHaloTree} \citep{Springel2005,DeLuciaBlaizot2007}. In Level 4, we have a total of 128 snapshots, and the maximum spacing between snapshots is 167 Myr which decreases towards higher redshift. In Level 3, we have a total of 64 snapshots, and the the maximum spacing between snapshots is 372 Myr which also decreases towards higher redshift. 

\begin{figure}
	\includegraphics[width=.41\textwidth]{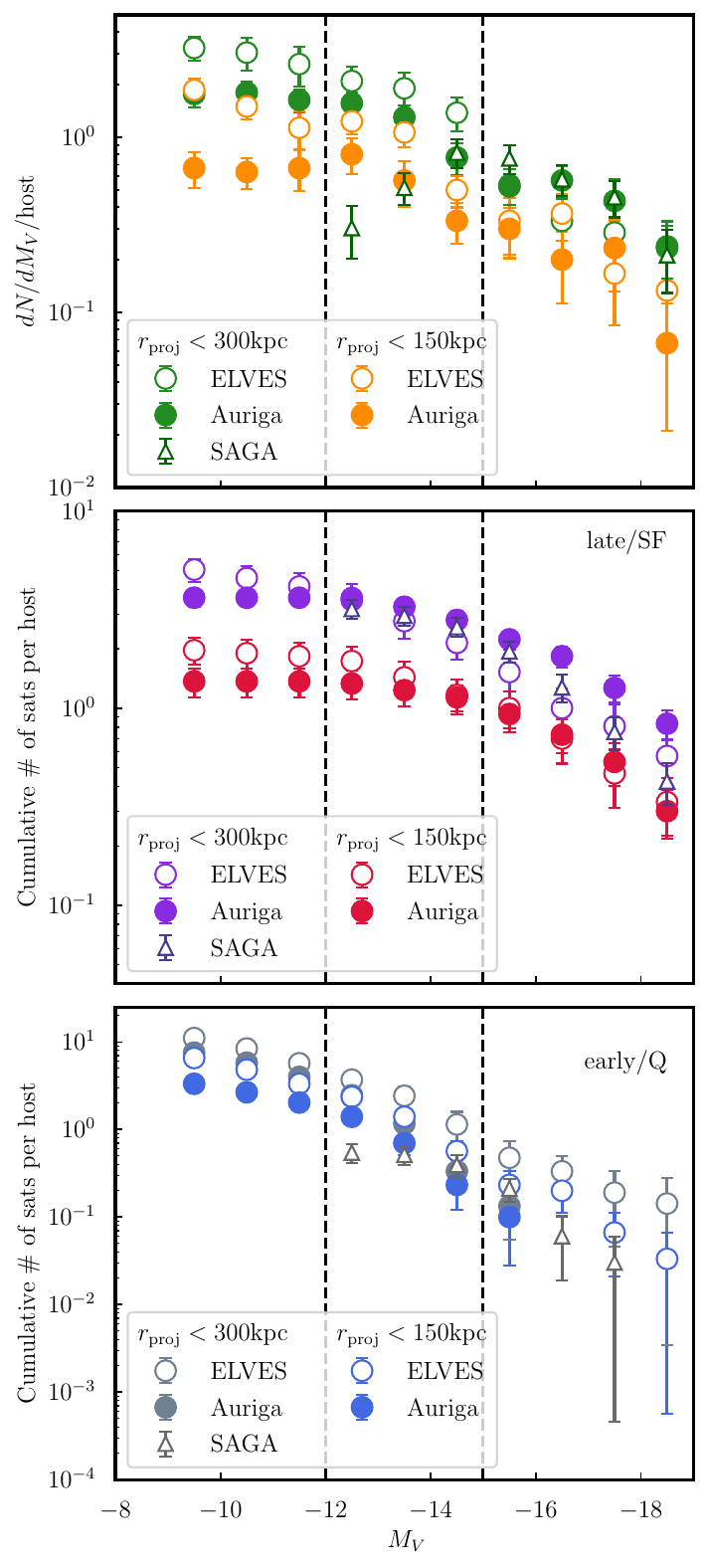}
    \caption{The satellite luminosity function (LF) in the Auriga simulations and  SAGA and ELVES sample. We only include SAGA satellites brighter than $M_V = -11.9$ since this is the magnitude limit cut. \emph{Top}: Differential luminosity functions for observed satellite galaxies in ELVES (open circles), SAGA (open dark green triangles), and simulated satellites in Auriga (solid green circles) are shown. The green (orange) points show LF of satellites within 300 kpc (150 kpc) of the host centre. In Auriga this is a 3D radial cut, whereas in ELVES this is a line-of-sight 2D projection radial cut. \emph{Middle}: Cumulative LFs for late-type satellite galaxies in the ELVES sample and star-forming satellite galaxies in and SAGA and Auriga within 150 and 300 kpc from the galaxy centre shown by violet, dark blue, and red points, respectively. \emph{Bottom}: Cumulative LFs for early-type satellite galaxies in the ELVES sample and quenched satellite galaxies in SAGA and Auriga within 150 and 300 kpc shown by grey, dark grey, and blue points, respectively. Overall, the shape of LFs in both simulation and observation is similar for all three cases, but ELVES tends to have more faint satellites than Auriga. Within the magnitude cut of $-15 \lesssim M_V \lesssim -12$ shown as the vertical dashed lines, the Auriga LF is in reasonable agreement with that in the ELVES sample for both late- and early-type galaxies. Auriga does not have bright quenched satellites within the radial cut of 150 kpc. The SAGA differential LF drops around $M_V \sim -12$ compared to both ELVES and Auriga, primarily due to a drop of quenched satellites shown in the bottom panel. The errobars show the error in the mean number of satellites per host within each magnitude bin.}
    \label{fig:LF}
\end{figure}

\begin{figure}
	\includegraphics[width=\columnwidth]{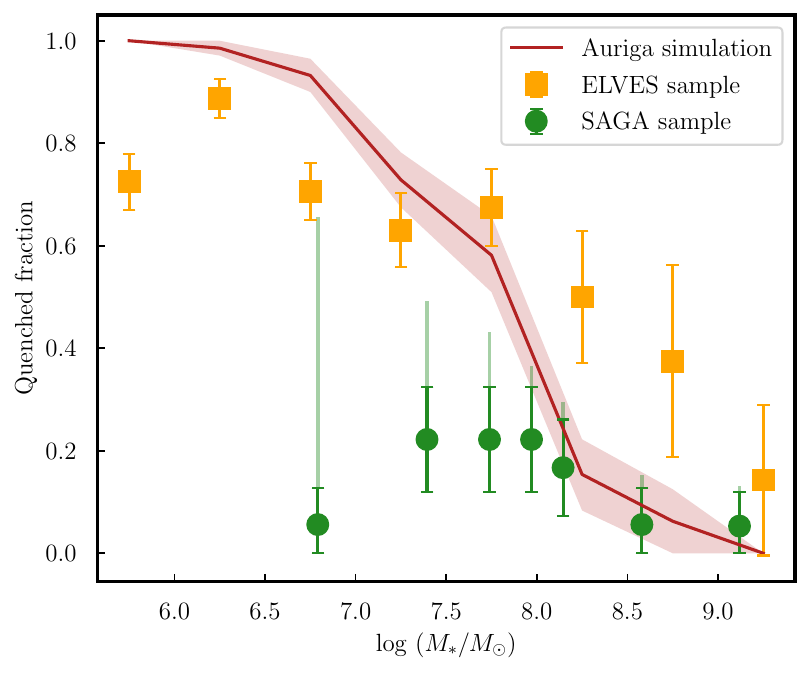}  
	\caption{Quenched fraction as a function of log stellar mass for ELVES, SAGA surveys and the Auriga simulations. Quenched fraction within each mass bin is defined as the number of quenched satellites divided by the total number of satellites in this mass bin. In the SAGA survey, a satellite galaxy is defined as quenched if it has no H$\alpha$ emission, while in the ELVES survey, it is defined by visually classifying it as early-type. In the Auriga simulations, it is defined as its gas phase ${\rm SFR} = 0$ at the end of the simulation. We apply the redshift incompleteness in \citet{Mao2021} to the SAGA quenched fraction and plot them as light green bars. We plot the error in the mean quenched fraction within each mass bin as errorbars (SAGA, ELVES) and a red-shaded region (Auriga).} \label{fig:fquenched}
\end{figure}

\begin{figure*}
	\includegraphics[width=0.9\textwidth]{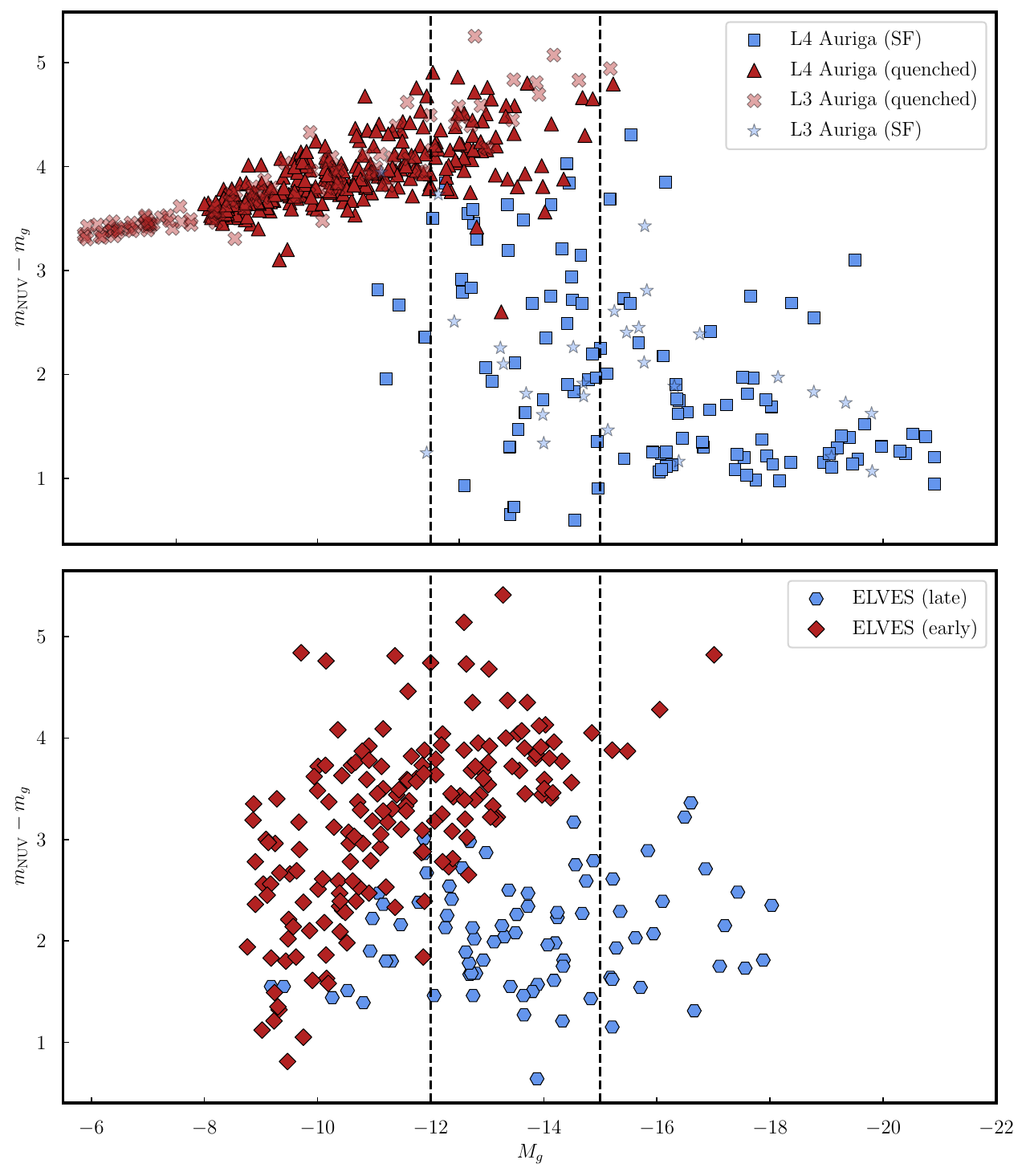}
    \caption{The colour-magnitude diagram for satellites in the ELVES sample and Auriga simulation selected within 300 kpc of the host centres. Top (bottom) panel shows the CMD for Auriga (ELVES) satellites. The Auriga satellites are separated into quenched (red triangles) and star-forming (blue squares), while the ELVES satellites  are separated into early-type (red open diamonds) and late-type (blue open hexagons). The two vertical dashed lines are the magnitude cut $-15 \lesssim M_g \lesssim -12$. Overall the distribution of model galaxies is similar in the magnitude range $-15 < M_g < -12$. There are fewer blue, faint satellite galaxies in the Auriga sample compared to the ELVES sample. We also include satellites from the higher resolution Auriga Level 3 simulation, denoted as stars and crosses, which probe fainter $(M_g > -8)$ regime.}
    \label{fig:CMD}
\end{figure*}

\subsection{NUV-g colours of Auriga satellites}
\label{sec:NUV_g_colour}

The absolute magnitudes in the $U$, $B$, $V$, $R$, $g$, $r$, $i$, and $z$ bands computed using models of \citet{BruzualCharlot2003} are provided for each halo and subhalo in the Auriga catalogs \citep{Grand2017}. To facilitate comparison of the colour distribution with observations of \citet{Carlsten2022} we also computed GALEX near-UV (NUV) magnitudes for each satellite using the Flexible Stellar Population Synthesis (FSPS) code \citep{Conroy2009,Conroy2010}\footnote{\href{https://github.com/cconroy20/fsps}{\tt https://github.com/cconroy20/fsps}} with its Python bindings \textsc{PyFSPS}\footnote{\href{https://github.com/dfm/python-fsps}{\tt https://github.com/dfm/python-fsps}}. Specifically, for each surviving satellite at $z=0$, we treated the satellite’s stellar particles (with known metallicity, age and mass) as a single-age stellar population and combined their individual luminosity to get total satellite luminosity.

We use the default \texttt{MIST} isochrones \citep{Dotter2016, Choi2016}, Chabrier IMF \citep{Chabrier2003}, and \texttt{MILES} spectral library \citep{Vazdekis2010, Falcon-Barroso2011} in FSPS, where the solar metallicity is 0.0142. However, creating a different SSP for each star particle is computationally expensive, since we have $\sim 5\times10^7$ particles in all the satellites. Instead, we use a method similar to that used by the Auriga SUBFIND code to compute the luminosity for each star particle by taking the a 2-D grid of metallicity and age, apply the grid to FSPS \texttt{get\_mags} function to compute the SDSS $g-$band and GALEX-NUV band magnitudes for each grid value, then interpolate all the star particles magnitudes based on this magnitude grid. We use the same age-metallicity grid used by Auriga's SUBFIND code.  We then convert the magnitudes into luminosity, add up all the particle luminosities in one satellite, and finally convert back to magnitudes and add an offset related to the particle mass to calculate the absolute satellite magnitude.

As an additional check, we computed the difference between the satellite absolute $g-$band magnitude we calculated and the satellite absolute $g-$band magnitude tabulated in the Auriga catalog as a function of the satellite $g-$band magnitude in the catalogue. We apply two different 3D radial cuts: $r < 300$ kpc and $r < 150$ kpc. We compute the 3D distance between the satellite and the host and apply a constant radial cut (300 kpc or 150 kpc) at all lookback times to select satellites. No matter what the radial cut is, the discrepancy between our value and the catalog value is less than 0.1 mag, indicating that our NUV and $g-$band magnitude calculation is robust and consistent with the catalogue values.

\begin{figure*}
	\includegraphics[width=\textwidth]{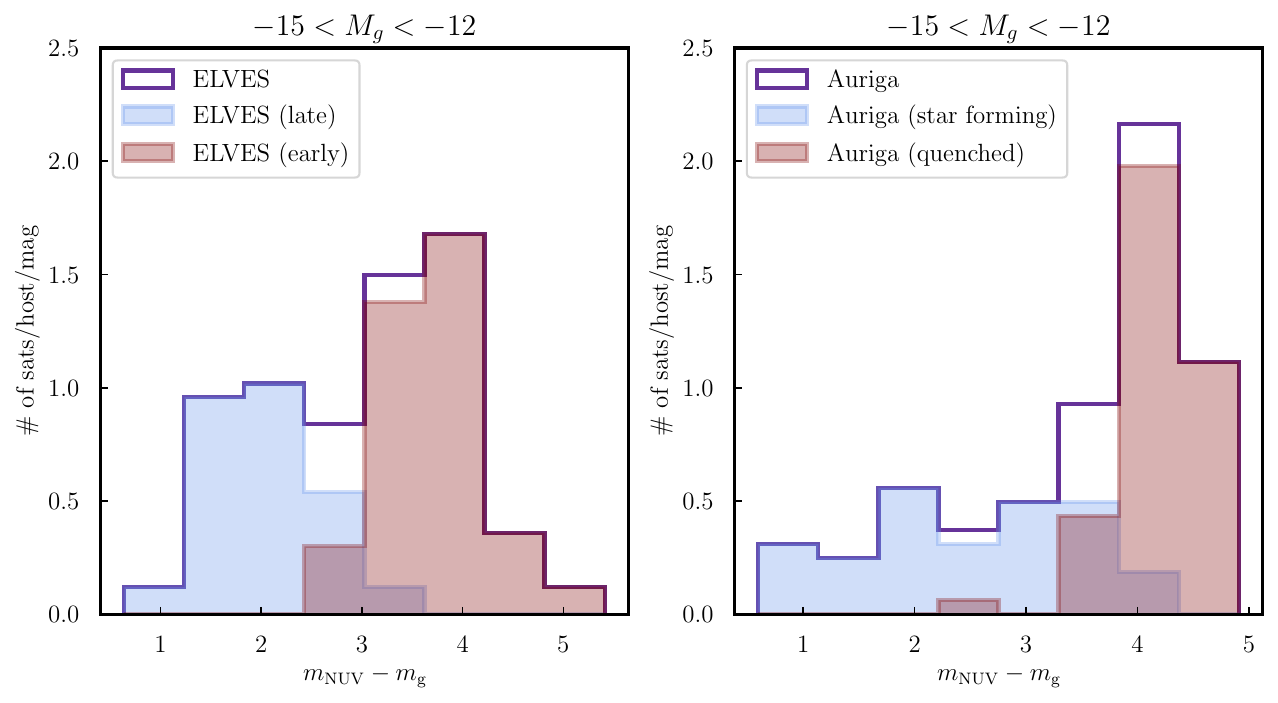}
    \caption{The distribution of NUV-$g$ colours in the Auriga and ELVES samples in the magnitude range $-15 < M_g < -12$. \emph{Left}: The distribution of NUV-$g$ colour of satellites in the ELVES sample (purple histogram) and for the early (red) and late type (blue) satellites. The $p-$value of the hypothesis that blue and red histograms are drawn from the same distribution is $< 10^{-5}$, indicating that the two populations are intrinsically different. \emph{Right}: The distribution of NUV-$g$ colour for the Auriga satellites (purple histogram), and for star forming (blue) and quenched (red) satellites separately.  The $p-$value in this case is also $< 10^{-5}$ and thus star forming and quenched satellites in the Auriga are similarly different to observed systems. The red peak in Auriga is somewhat redder, while the blue peak is broader and lower than for the ELVES satellites. The $p-$value for the ELVES and Auriga colour distributions is $8\times10^{-4}$, so we cannot rule out the possibility that they are different. Note that we do not show that the colour distribution is bimodal - we simply show that there are two distinct populations in the colour distribution.}
    \label{fig:colour_bimodal}
\end{figure*}

\begin{figure*}
	\includegraphics[width=\textwidth]{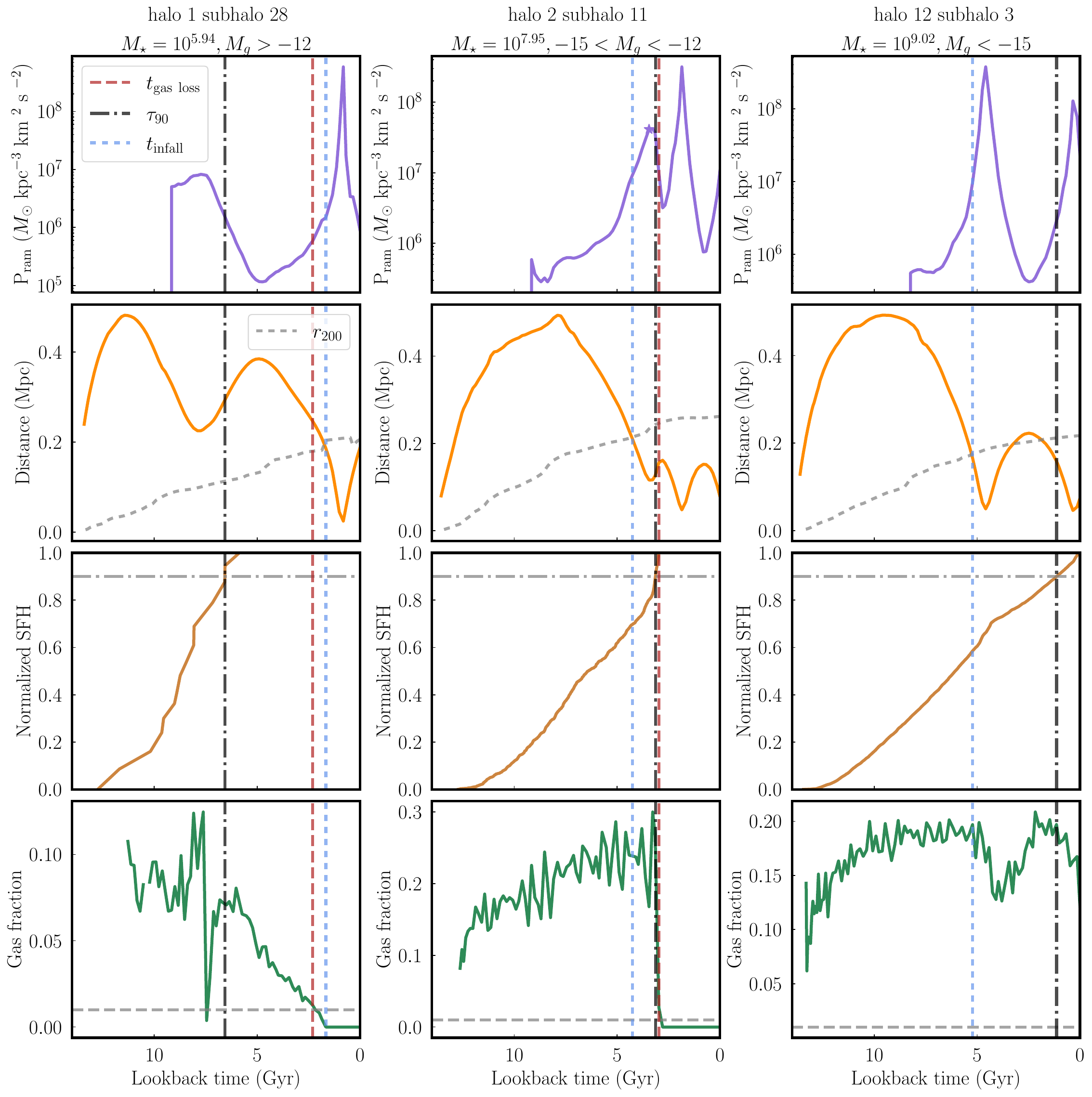}    \caption{An illustration of ram pressure (top), distance from the center of the host (second row from top), star formation history (SFH, 3rd row from the top) and gas fraction (bottom) as a function of lookback time for three satellites in three different magnitude bins: $M_g > -12$, $-15 < M_g < -12$, $M_g > -15$. The dashed lines  overplot three timescales, $t_{\mathrm{gas loss}}, t_{\mathrm{infall}}, \tau_{90}$ as red, blue, and black dashed lines, respectively. The black dashed line in the distance row shows the evolution of the host $r_{200}$ (the radius within which the halo's mean density is 200 times greater than the critical density of the universe). A horizontal dot-dashed grey line in the normalized SFH and gas fraction rows show a SFH value of 0.9 and a gas fraction value of 0.01, respectively.
	}
    \label{fig:rp_illustration}
\end{figure*}

\begin{figure*}
	\includegraphics[width=\textwidth]{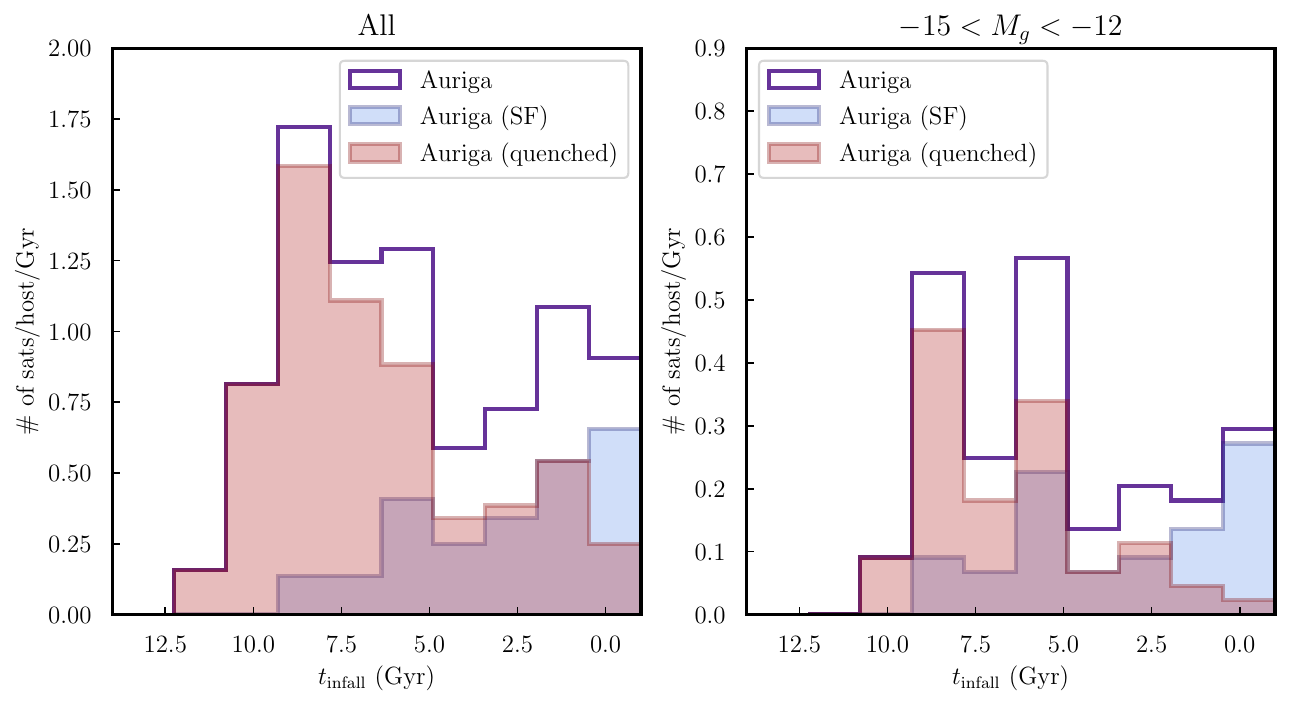}
    \caption{The distribution of lookback infall times in the Auriga simulations \emph{Left}: Distribution of lookback infall times per host per Gyr for all satellites within 300 kpc of the galaxy at $z=0$ (purple line) in the Auriga simulation and for the star-forming (blue) and quenched satellites (red). The $p-$value for the red and blue histograms to be drawn from the same distribution  $<10^{-5}$, indicating that the two populations are intrinsically different. \emph{Right}: The same distribution, but for all satellites within the magnitude cut $-15 \lesssim M_V \lesssim -12$ for a direct comparison with the ELVES data. The $p-$value in this case is $1.5\times 10^{-3}$. Here again quenched satellites tend to have earlier infall times than star-forming ones, in agreement with Figure 2 in \citet{Fillingham2019}.}
    \label{fig:infall_bimodal}
\end{figure*}

\begin{figure*}
	\includegraphics[width=0.8\textwidth]{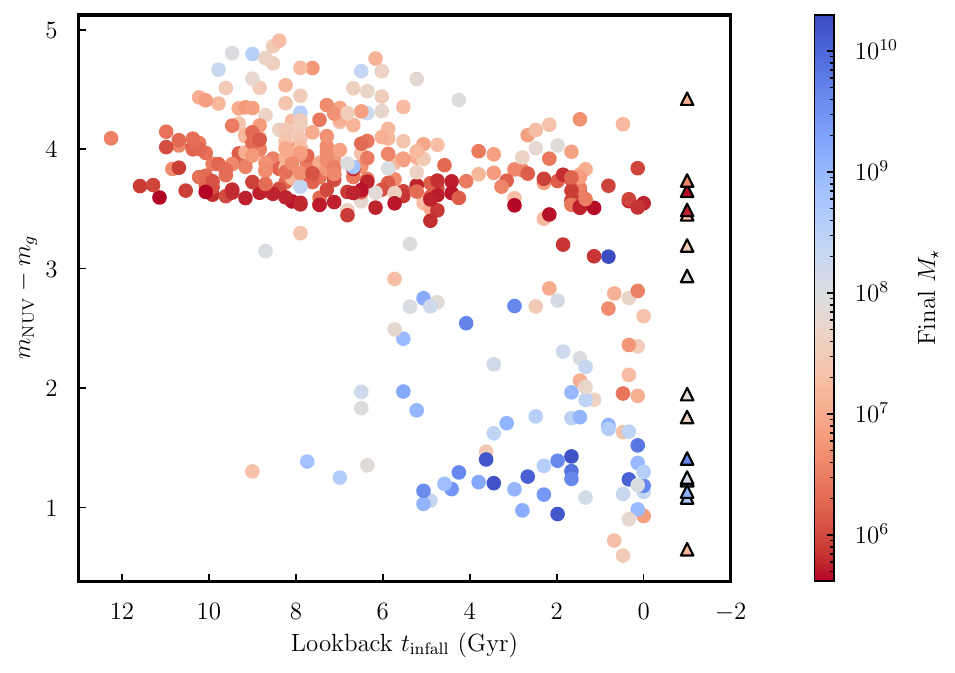}
    \caption{NUV-$g$ colour vs $t_{\rm{infall}}$ color coded by final $M_\star$ for all satellites in L4 that are within 300 kpc of their host by the end of the simulation. This panel connects the colour distribution with the infall time distribution. Triangles with a black frame represent satellites that are within 300 kpc of their host by the end of the simulation but outside of the virial radius, thus they are never environmentally influenced by their host. We denote their $t_{\rm{infall}} = -1$ Gyr.}
    \label{fig:color_infall}
\end{figure*}


\section{Observable properties of simulated satellites}
\label{sec:comparison} 
In this section, we present the luminosity function, the colour distributions, and the colour--magnitude diagram for satellites in the Auriga simulations. We also compare these trends the ELVES and SAGA observational surveys.

\subsection{Luminosity function}
\label{sec:LF}

We first compare the differential and cumulative luminosity functions (LFs) of satellites in the Auriga simulation and ELVES and SAGA sample within radial cuts of $r_{\mathrm{proj}} < 300$ kpc and $r_{\mathrm{proj}} < 150 $ kpc. Note that the radial cut in the Auriga simulations is 3D, whereas the radial cut in ELVES and SAGA samples is a 2D line-of-sight projection. There are 21 ELVES galaxies surveyed out to 300 kpc, and 30 hosts surveyed out to 150 kpc. 33 hosts are surveyed out to 300 kpc in SAGA. We apply the same radial cuts to the 30 hosts in the Auriga Level 4 halos, and compare the differential and cumulative $V$-band LF of the simulated satellites with the LF of ELVES and SAGA satellites. Note that the high-resolution zoom region around each Auriga host extends to $\sim$1 Mpc, well beyond the radial extent of the ELVES and SAGA satellite sample. 

The comparison is shown in Figure~\ref{fig:LF}. 
The two vertical dashed lines in Figure~\ref{fig:LF} mark the magnitude range of $-15 < M_V < -12$. \citet{Carlsten2022} found that satellites in the ELVES sample transition in this magnitude range from being mostly late type to mostly early type.
We thus also focus on satellites in this magnitude range for comparison with observations and to understand the mechanisms that drive this shift in properties. 
We define star-forming satellites in the simulations as having a star formation rate SFR $> 0$ at $z = 0$. Late-type ELVES satellites (which are more likely to be star forming) are defined by visually inspection of their morphology \citep[see Section 6.2 of ][for a detailed discussion]{Carlsten2022}. In SAGA, quenched satellites are defined as having EW(H$\alpha) < 2\angstrom$.

We count the number of satellites in each magnitude bin for each sample we explore. Following \citet{Carlsten2022}, we quantify the variation in the luminosity function between hosts by computing the standard deviation of satellite abundance across different hosts in both Auriga, SAGA and ELVES and dividing it by $\sqrt{N_{\mathrm{host}}}$. Thus, the error bars shown in Figure~\ref{fig:LF} denote the intrinsic host-to-host scatter in satellite abundance. 

The top panel in Figure~\ref{fig:LF} shows that there are somewhat more satellites per host per magnitude bin in the Auriga simulations than in the ELVES and SAGA sample on the bright end of LF, but both are consistent within the errorbars. This trend is reversed as we proceed to fainter satellites. The middle and bottom panels showing differential LFs indicate that this trend is a combination of different trends for the star-forming and quenched galaxies. LFs of star-forming galaxies in the simulation are quite close to observed ones, especially in the luminosity range $-15 < M_V < -12$. The LF shapes are somewhat different, however. The shapes of LFs of quenched/early type galaxies, on the other hand, are quite similar but the amplitude of the simulated LFs is $\sim 1.3-2$ times lower than in the ELVES sample. We found that this lack of satellites in Auriga below $M_V = -12$ is not due to resolution limits by confirming the convergence of L3 and L4 LF down to $M_V \sim -8$.  We also note that \citet{Grand2021} shows convergence of the model for a single host halo down to $M_V \sim -8$ with resolution greater than L3. Notably, the transition from the mostly quenched to mostly star-forming satellites in both the Auriga simulations and ELVES and SAGA sample occurs in the range $-15 < M_V < -12$.

\subsection{Quenched fraction}

In Figure~\ref{fig:fquenched} we compare the satellite quenched fraction as a function of the logarithm of stellar mass in the ELVES and SAGA surveys to the Auriga simulations. We apply a 3D radial cut of 300 kpc for the Auriga sample and a 2D line-of-sight radial cut of 300 kpc for the ELVES and SAGA sample. In the Auriga simulations, we define a satellite as quenched if its gas phase ${\rm SFR} = 0\ M_{\odot}$yr$^{-1}$ at the end of the simulation. \citet{Karunakaran2021} investigated other definitions of SFR such as the average mass of star particles formed over the last gigayear and found that both estimates of SFR produce similar results in terms of the quenching status of satellite galaxies. Moreover, by analyzing the left panel of Figure 2 in \citet{Karunakaran2021}, the specific SFR (sSFR) values for most star-forming satellite galaxies (SFR > 0 $M_{\odot}$yr$^{-1}$) are above 0.01. Therefore, a threshold of sSFR = 0.01 to distinguish quenched and star-forming satellite galaxies would produce similar results in our analysis. In the ELVES survey, we use morphology-based classification of quenched satellites of \citet[][]{Carlsten2022}, while SAGA satellites are classified using H$\alpha$ emission by \citet[][]{Mao2021}. All the errorbars are estimated using a bootstrap sampling strategy. 

Figure~\ref{fig:fquenched} shows that although the quenched fraction in the SAGA survey is significantly smaller than that in the Auriga simulation for satellites with $M_\star\lesssim 10^8\, M_\odot$, the new ELVES survey has a much higher quenched fraction for such satellites and is much closer to the quenched fraction measured in the Auriga simulations. For satellites with $M_\star\gtrsim 10^8\, M_\odot$, the quenched fraction in simulations matches that in the SAGA survey and is smaller than the fraction in the ELVES survey, but consistent at the 2-sigma level. On the lower-mass end, the Auriga simulations is inconsistent with both ELVES and SAGA but the discrepancy with ELVES is less pronounced. A more recent study by \citet{Karunakaran.etal.2022} shows that the discrepancy between SAGA and ELVES could potentially be significantly reduced by using a consistently-derived sSFR
and absolute magnitude limit in both samples.

\subsection{Colour--magnitude diagram}
\label{sec:CMD}

Figure~\ref{fig:CMD} shows the NUV-$g$ colour as a function of the $g$-band absolute magnitude $M_g$ for satellites in the ELVES and Auriga Level 3 and Level 4 simulations. It shows that ELVES and Auriga satellites basically occupy the same region in the NUV-$g$ and $M_g$ parameter space in the magnitude range $-15 < M_g < -12$, but the simulated satellite population is deficient in blue, faint satellites in the lower left of this parameter space that are present in ELVES. We discuss in detail possible reasons for this in Section~\ref{sec:compare_obs}.

This figure shows that there is a transition from early-type or quenched satellites to late-type or star-forming within the magnitude range $-15 < M_g < -12,$ for both Auriga and ELVES satellites. Satellites that are brighter than $M_g \sim -15$ are primarily star-forming or late-type in both Auriga and ELVES. 

We also include Auriga Level 3 satellites in the colour magnitude diagram. Due to 8 times better resolution compared to Level 4, satellites in the Level 3 simulation reach much fainter satellites ($M_g \gtrsim -8$). Nevertheless, the improvement in resolution does not remedy the discrepancy between the simulations and observations caused by dearth of blue, faint satellites in the former.

To examine this transition further, Figure~\ref{fig:colour_bimodal} shows the distribution of NUV-$g$ colour in the ELVES and Auriga L4 satellites. We separate ELVES satellites into late- (blue shaded histogram) and early-types (red-shaded histograms), and Auriga satellites into quenched (red-shaded histograms) and star-forming (blue shaded histogram). The $p-$value for the red and blue histograms to be drawn from the same parent distribution for both ELVES and Auriga satellites is $<10^{-5}$, indicating a statistically significant difference. Thus, for both observation and simulations, we find two distinct populations with respect to satellite colour. 

Overall, results presented here show that the Auriga simulations capture certain qualitative trends such as luminosity function, quenched fraction, and the two distinct populations in terms of NUV$-g$ colour that are also found in observed samples in ELVES and SAGA. We now turn to exploring the physical processes that drive satellite quenching in the simulations. We will return to discussing the comparisons between simulations and observations and what that means for the Auriga model in Section~\ref{sec:discussions}.

\begin{center}
\begin{figure}
	\includegraphics[width=0.85\columnwidth]{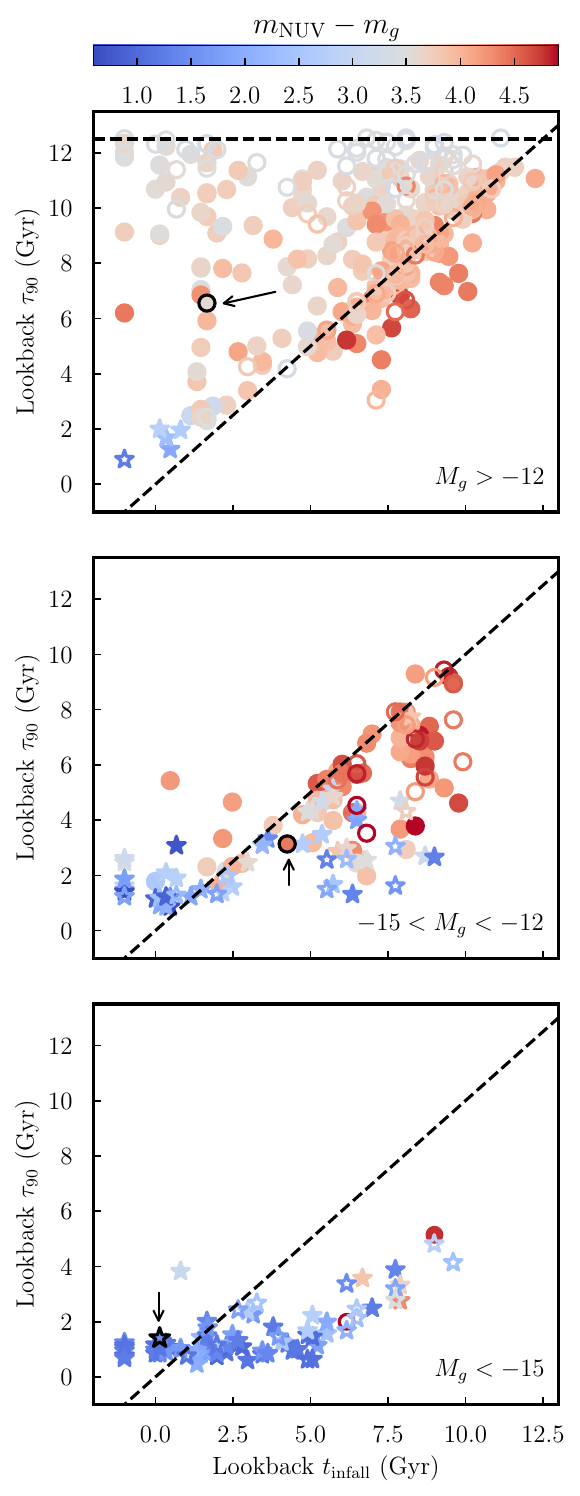}
    \caption{$\tau_{90}$ as a function of $t_{\mathrm{infall}}$. Satellite points in all panels are colour-coded by their NUV-$g$ colour shown in the colour bar. Different rows show satellites in three different magnitude bins. Open (solid) symbols refer to the satellites in Level 3 (Level 4) simulation. Star symbols refer to satellites that are star-forming, while circles refer to quenched satellites in our definition. The three examples in different magnitude bins in Figure ~\ref{fig:rp_illustration} are indicated by black arrows and a black frame. 
    Satellites that never cross the host's virial radius but are within 300 kpc by the end of the simulation are assigned the infall time of $t_{\mathrm{infall}} = -1$. The diagonal dashed line delineates positive and negative $t_{\mathrm{delay}}$: above this line, nearly all satellites are faint, low-mass and quenched before infall, whereas below this line the satellites have  stellar masses of $M_\star \gtrsim 10^7 M_{\odot}$ and are more likely to experience environmental quenching effects. The horizontal dashed line in the upper right panel corresponds to a lookback time of 12.5 Gyr ($z \approx 6$), the end of the epoch of reionization in the simulation.}
    \label{fig:tau90_tinfall}
\end{figure}
\end{center}

\begin{center}
\begin{figure*}
	\includegraphics[width=\textwidth]{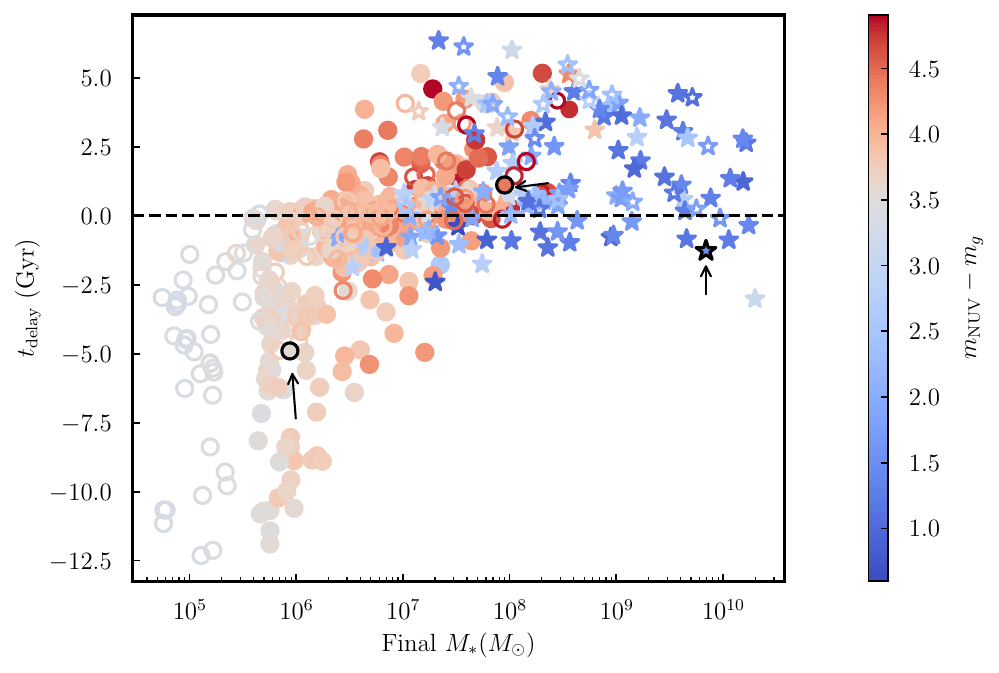}
    \caption{Delay time as a function of final stellar mass. Symbols are the same as in Figure~\ref{fig:tau90_tinfall}.
    For quenched satellite galaxies, a negative $t_{\mathrm{delay}}$ means they are quenched before infall; here these are primarily satellites with $M_\star \lesssim 10^6 M_{\odot}$ in the lower left. }
    \label{fig:tdelay_mass}
\end{figure*}
\end{center}

\begin{figure*}
	\includegraphics[width=\textwidth]{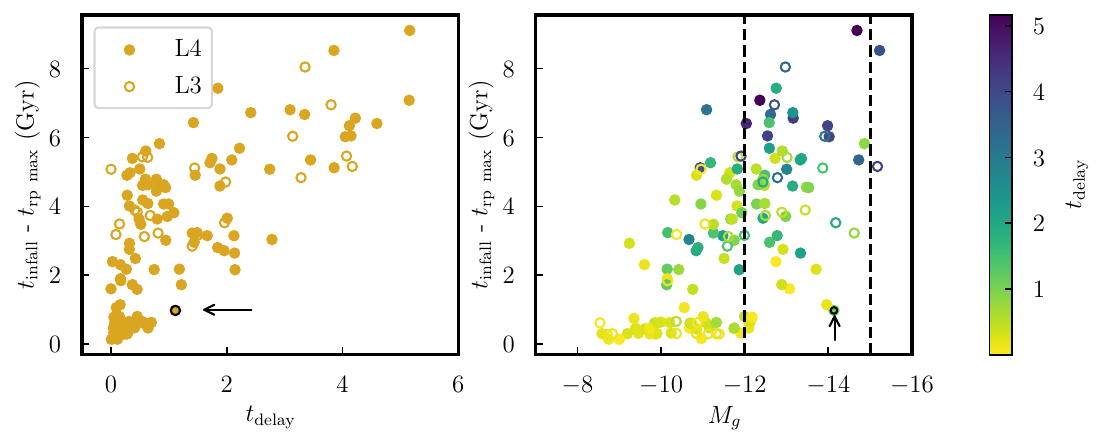}    
	\caption{{\emph{Left}: Time interval between infall and local ram pressure peak after infall as a function of \tdelay, for all quenched satellites at $z=0$ that lose their gas after infall. Halo 2 subhalo 11 in Figure~\ref{fig:rp_illustration} that loses its gas after infall is indicated by a black arrow. We see a correlation that the higher \tdelay in general has higher $t_{\rm{infall}} - t_{\rm{rp\ max}}$, albeit with some scatter}. \emph{Right}: Time interval between infall and local ram pressure peak after infall as a function of the $g$-band absolute magnitude, $M_g$, colour coded by \tdelay, for all quenched satellites at $z=0$ that lose their gas after infall. The colour bar is set such that the colour divergence starts at \tdelay = 0, delineating positive and negative \tdelay. Halo 2 subhalo 11 is indicated by a black arrow. We do not include subhalo 2 and subhalo 28 of halo 1 and because the former is still star-forming at $z=0$ and the latter is quenched before infall. Open (solid) symbols refer to Level 3 (Level 4) satellites.  Lower mass satellites  with $M_\star \leqslant 10^6 M_{\odot}$ generally reach a local ram pressure peak immediately after infall, indicating that ram pressure stripping is their main quenching process. They also have a negative or close to 0 \tdelay. Some of the intermediate mass satellites, $10^6 M_{\odot} \leqslant M_\star \leqslant 10^8 M_{\odot}$) are also immediately quenched by ram pressure after infall, but a fraction is quenched on much longer time scales after infall, consistent with a positive \tdelay. The most massive satellites ($M_\star \gtrsim 10^8 M_{\odot}$) tend to reach the  ram pressure peak on much longer time scales after infall. }
    \label{fig:trampeak}
\end{figure*}


\section{Physical evolution time scales and star formation quenching}
\label{sec:timescales}

Given that luminosity function, quenched fraction, and the two distinct populations in terms of NUV$-g$ colour in the Auriga simulations matches the distribution of observed galaxies in the ELVES sample reasonably well, we can use simulations to gain insight into the quenching 
processes that separate these two types of satellites. To this end, we consider five different characteristic epochs and time scales associated with evolution of satellites in simulations: 
\begin{itemize}
\item[(i)] The quenching time $\tau_{90}$ is defined as the lookback time when $90\%$ of the satellite stars have formed,\footnote{\citet{Weisz2015} found that using $\tau_{90}$ instead of $\tau_{100}$ could potentially minimize the uncertainty induced by modeling blue straggler populations. For star-forming satellite galaxies, this quantity does not indicate their quenching status; rather, it is an upper bound of the lookback quenching time.} 

\item[(ii)]
The gas loss time, $t_{\rm{gas\ loss}}$, is defined as the lookback time when the gas mass fraction (the ratio of the gas mass to total mass) first drops below 0.01.

\item[(iii)] The infall time,  $t_{\mathrm{infall}}$, is the lookback time when the satellite first crosses the virial radius $r_{200}$. 

\item[(iv)] The time interval between  $t_{\mathrm{infall}}$ and $\tau_{90}$, $t_{\mathrm{delay}} = t_{\mathrm{infall}} -  \tau_{90}$, which gauges whether quenching occurs before or after satellite infall.  A positive \tdelay means the system stopped forming stars after infall and a negative value means it stopped before infall.

\item[(v)] The lookback time when the ram pressure experienced by gas in the satellite reaches a local maximum after infall \emph{and} is closest to $t_{\mathrm{gas\ loss}}$, $t_{\rm rpmax}$. We only compute this timescale for satellites quenched after infall to make sure we are only including satellites for which environmental processes are relevant. We require the local ram pressure maximum to be closest to $t_{\mathrm{gas\ loss}}$ to make sure that we are finding the most relevant pericentre passage that quenches the satellite. 
\end{itemize}

To calculate the ram pressure, we use the scaling of the ram pressure force with background gas density, $\rho_{\rm CGM}$, and relative satellite velocity, $v_{\rm sat}$, that follows from dimensional considerations: 
\begin{equation}
    P_{\mathrm{ram}} =\rho_{\mathrm{CGM}} v^2_{\mathrm{sat}}.
\end{equation}
\label{ram_pressure_equ}
We estimate $\rho_{\rm CGM}$ using  spherically averaged density profile extending out to $4r_{200}$ around each host in each time snapshot, interpolating the density profiles at the subhalo's position to get $\rho_{\mathrm{CGM}}$ at a specific epoch between snapshots \citep[see][for a detailed discussion of this approach to computing $P_{\rm ram}$]{Simpson2018}. This calculation of $\rho_{\mathrm{CGM}}$ may be biased in two ways. First, in some cases, the host halo density profile is not homogeneous. Second, there is a certain error in estimating the pericentric passage time  due to finite spacing between simulation snapshots. This can potentially be improved using orbit integration method to interpolate between snapshots \citep{Richings2020}, but we postpone exploration of such methods to future work.

As an illustration of how different timescales relate to satellite observables, Figure~\ref{fig:rp_illustration} shows evolution of ram pressure, distance to the host, star formation, and gas fraction for three satellites of different $M_g$ and stellar mass.
The faintest satellite is quenched before infall and its star formation is halted at an early times ($\sim$ 6 Gyr). The intermediate luminosity satellite is quenched after infall, likely by environmental effects such as ram pressure stripping. The local ram pressure peak that we identify as being the most relevant in quenching this satellite is marked as a purple star. Although the second later ram pressure peak has a larger amplitude, it is the first peak that strips all of the gas as can be see in the gas fraction panel. There is no gas left at the time of this second ram pressure peak. The luminous satellite is still forming stars at $z=0$. Although it experienced two local ram pressure peaks after infall, it is massive enough to resist these ram pressure events.

\subsection{Satellite infall time distribution}
\label{sec:infall_distribution}

\citet{Simpson2018} found indications that satellite infall times in the Auriga simulations are bimodal (see their Figure 11). Here, we re-investigate the infall time distribution using the optimal histogram bin width estimated with the objective Bayesian method of \citet{Knuth.2006} for each sample we consider in Figure~\ref{fig:infall_bimodal}.

We apply the Kolmogorov-Smirnov (KS) statistics to the infall time distributions of blue and red distributions in both panels to quantify the $p$-value that the two populations are drawn from the same parent distribution and find that it is less than $10^{-5}$. This indicates that these distributions statistically distinct and thus in agreement with conclusions of \citet{Simpson2018} that distributions of infall times of star-forming and quenched satellites are staistically different. This, in turn, 
implies that environmental effects play an important role for at least a substantial fraction of satellites. 

In Figure~\ref{fig:color_infall}, we combine Figure~\ref{fig:colour_bimodal} and Figure~\ref{fig:infall_bimodal} to directly probe the relation between NUV-$g$ colour and $t_{\rm{infall}}$. There are two obvious groups of points: one in the upper left with early $t_{\rm{infall}}$ and red colour (NUV-$g \geq 3.5)$, and the other in the lower right with late $t_{\rm{infall}} (t_{\rm{infall}} \leq 6$ Gyr) and blue colour (NUV-$g \leq 3)$. This indicates that early infall satellites tend to be red and quenched, while late infall satellites are blue and star-forming. There are some recent infall satellites with red colour (upper right), but they have low stellar mass $(M_\star \leq 10^7 M_\odot)$, indicating that they are not able to retain their gas after interacting with their hosts and get quenched on short timescales. Dwarfs that never fall within the virial radius of their host (black triangles) display a range of colour, but most of them have blue colours. Studies have shown that field dwarfs tend to be bluer than their satellite counterparts \citep{Geha2012}. Although we do not have a statistical sample of real dwarfs here, the population of satellites that never fall within the virial radius of their host does demonstrate a slight trend of bluer colour.

\subsection{How $t_{\rm{infall}}$ relates to $\tau_{90}$}
\label{sec:tau90_tinfall}
The panels of Figure~\ref{fig:tau90_tinfall} show distribution of satellite galaxies in three absolute magnitude ranges in the $t_{\rm infall}-\tau_{90}$ plane. The upper panel of Figure~\ref{fig:tau90_tinfall} shows that most faint satellites with $M_g > -12$ are quenched before they became satellites (i.e., have
$\tau_{90}<t_{\rm infall}$). There is a group of low-luminosity satellites in Level 3 (marked by open symbols in both Figures~\ref{fig:tau90_tinfall} and ~\ref{fig:tdelay_mass}) that have a lookback $\tau_{90}$ in the range $\approx 10-12.5$ Gyrs ($z\approx 2-6$), after the end of the epoch of reionization in the Auriga simulation at $z$ = 6 \citep{Grand2017, Vogelsberger2013}. We only have such low-mass systems in Level 3 simulations because these systems are not resolved in the Level 4 simulations. The quenching mechanism here is most likely suppression of gas accretion due to UV heating after reionization. Indeed, previous studies found that reionization mainly affects satellites with a stellar mass $M_\star \lesssim 10^6 M_{\odot}$ \citep[e.g.,][]{Bose.etal.2018,Kravtsov.Manwadkar.2022}. 

The middle panel of Figure~\ref{fig:tau90_tinfall} shows a sharp transition in the $t_{\rm{infall}} - \tau_{90}$ plane: satellites in the magnitude range $-15 < M_g < -12$ are either still star-forming or quenched by environmental processes since they have $t_{\rm{infall}} > \tau_{90}$. The bottom panel of Figure~\ref{fig:tau90_tinfall} shows that luminous satellites $(M_g < -15)$ are predominantly star-forming, and they became satellites less than $\approx 7$ Gyrs ago (i.e. after $z\sim 1$). Indeed, there are some massive satellites with $t_{\rm infall}<2$ Gyrs in the lower-left region of the bottom panel that are actively star-forming, similar to Large and Small Magellanic Clouds. The lack of quenched massive satellites with early $t_{\rm infall}$ is likely due to their disruption or merging with the central galaxy.

Figure~\ref{fig:tdelay_mass} also shows that a substantial fraction of satellites with stellar masses in the range of $M_\star \lesssim 10^7\, M_{\odot}$ are quenched before infall (negative $t_{\mathrm{delay}}$), although this fraction decreases with increasing $M_\star$. This quenching is thus likely due to internal processes such as stellar feedback, which can drive out most of the gas and quench dwarf galaxy for an extended period of time \citep[see, e.g.,][]{Rey.etal.2022}, or a combination of gas suppression due to UV heating and internal feedback \citep{Rey.etal.2020}, or interactions with the filaments of the cosmic web \citep{Benitez.Llambay2013}.

In contrast with small-mass galaxies with $M_\star<10^7\, M_\odot$ that are predominantly quenched and red, there are both star-forming and quenched galaxies among more massive satellites with the fraction of star-forming galaxies increasing with stellar mass. Figure~\ref{fig:tau90_tinfall} and ~\ref{fig:tdelay_mass} together show that quenched satellites with $M_\star\gtrsim 10^7\, M_\odot$ tend to have earlier infall times. The quenching of star formation in satellites of this mass range is thus predominantly due to environmental effects. Given that these effects operate on a certain timescale, satellites that accrete sufficiently early are quenched, while those that accrete late can remain star forming (we will discuss this further in the next section). 

Figure~\ref{fig:tdelay_mass} also shows that intermediate mass satellites $10^7 M_{\odot} \lesssim M_\star \lesssim 10^8 M_{\odot}$ quench at the time close to the infall epoch. Note that crossing $r_{200}$ ($t_{\rm infall}$) marks the onset of environmental effects only very approximately because galaxies can start to experience enhanced tidal and ram pressure forces well before crossing the virial radius \citep[e.g.,][]{Behroozi.etal.2014}. Thus, small negative $t_{\rm delay}$ can still be due to quenching by environmental processes.

\subsection{Timescales of environmental quenching}
\label{sec:ram_pressure}

To better understand the connections between ram pressure, infall, and observables such as mass and $M_g$, we plot the time duration between infall and the closest local ram pressure peak to $t_{\mathrm{gas loss}}$ as a function of satellite $M_g$ colour in Figure~\ref{fig:trampeak}. Symbols are the same as Figure~\ref{fig:tau90_tinfall} and ~\ref{fig:tdelay_mass}, but we {\it only} select satellites that are quenched by the end of the simulation and lose their gas after infall since we want to explore {\it only} the effects of environmental quenching.

In general, the left panel of Figure~\ref{fig:trampeak} shows that there is a correlation between {\tdelay}\, time and $t_{\rm{infall}} - t_{\rm{rp\ max}}$. Note that we choose $t_{\rm{rp\ max}}$ as the time closest to the quenching time $t_{\rm{gas\ loss}}$ to make sure that we are capturing the most relevant ram pressure stripping event that causes quenching. We shift both {\tdelay}\, and  $t_{\rm{rp\ max}}$  by $t_{\rm{infall}}$ to account for any infall time differences. This correlation shows that ram pressure stripping is one of the main quenching mechanisms, and the scatter indicates that it is not the only mechanisms at play.

The right panel of Figure~\ref{fig:trampeak} shows that faint satellites ($M_g > -12$) reach local ram pressure peak within
$\approx 0.5-1$ Gyr after infall and are thus likely quenched by ram pressure stripping of the first pericentric passage shortly after infall. This is also true for a fraction of satellites of intermediate luminosity ($-15 < M_g < -12$). However, a fraction of satellites within this luminosity range experiences peak ram pressure force $\sim 2-7$ Gyrs after infall, likely after multiple pericentric passages until quenched by a later ram pressure peak. The peak ram pressure force time of the luminous quenched satellites ($M_g < -12$)  has the broadest distribution, likely also due to quenching by starvation. 

By visually inspecting evolution of the ram pressure force, distance to the host halo centre history, star formation history, and gas fraction history for every satellite, we identify four possible reasons for the scatter of $t_{\rm{infall}} - t_{\rm{rp\ max}}$ at a fixed $M_g$ in Figure~\ref{fig:trampeak}. 

First, there are multiple instances where a satellite loses most of its gas between two peaks in ram pressure force, and it is difficult to conclude which peak is more relevant in quenching the satellite. It is possible that after the first pericentre passage a significant amount of gas is stripped, or it could also be that a significant amount of gas is stripped when the satellite is on its way to the second pericentric passage. 
Second, the estimation of $t_{\rm{rp\ max}}$ itself is uncertain, since some satellites have multiple local peaks that are close to each other around $t_{\mathrm{gas\  loss}}$.
Third, we use a spherically averaged density profile in computing $\rho_{\mathrm{CGM}}$, which is only an approximation to the true local density that the satellite experiences. Fourth, variations in satellite orbit and impact parameter could also drive a scatter. 

Overall, results presented in this section indicate that satellites of mass $M_\star \gtrsim 10^7 M_\odot$ are quenched mainly after they become satellites by ram pressure stripping. The quenched fraction in this mass range is thus determined by the fraction of satellites that had sufficiently early infall time and had sufficient time to experience significant gas stripping due to ram pressure stripping. 

In addition, results presented in this and previous sections indicate that the time scale of such stripping depends on satellite's stellar mass and details of its orbit. The smallest mass satellites are quenched shortly after infall, while larger mass satellites exhibit a broad range of quenching time scales depending on their orbit. This introduces scatter in the quenching time and colour of satellites with similar infall time. Most of the satellites with the largest stellar masses ($M_\star\gtrsim 10^8\, M_\odot$) are not quenched and continue to form stars to $z=0$. This implies that such massive satellites are resilient against typical ram pressure forces they experience during evolution. At the same time, massive satellites with earliest infall time that could be quenched may not survive to $z=0$ due to dynamical friction and associated tidal disruption and merging they experience. 

\section{DISCUSSION}
\label{sec:discussions}
\subsection{Comparison with observations}
\label{sec:compare_obs}
In recent years, larger and deeper observations have granted us an unprecedented sample of dwarf satellites around MW analogs, important probes of galaxy formation and the nature of dark matter. In particular, Satellites Around Galactic Analogs  \citep{Geha2017, Mao2021} contains classical bright satellites ($M_r < -12.3$ mag) of 100 MW-analogs in the distance range $20 < D < 40$ Mpc. There has been an ongoing tension between SAGA and simulations such as APOSTLE  \citep[A Project Of Simulating The Local Environment,][]{Sawala2016, Fattahi2016}, the DC Justice League simulations \citep{Applebaum.etal.2021}, ARTEMIS \citep{Font2020}, and Auriga, where SAGA found significantly more star-forming, low-mass satellites than these simulations \citep{Karunakaran2021,Akins.etal.2021}. Although the large discrepancy between simulations and observations in terms of satellite quenching fraction in the low-mass regime can be potentially mitigated by considering the differences in host mass distributions and observation selection effects \citep{Font2022}, the question of whether simulations can reproduce observations has persisted until the release of the ELVES Survey. 

Compared to SAGA, the quenched fraction of satellites in the ELVES sample is more consistent with the results from simulations. Here, we take the ELVES data and look into the colour distribution of satellites to understand the different quenching mechanisms and the timescales  at play. We found the NUV-$g$ colour distributions across Auriga and ELVES are similar, with a prominent red peak at around $m_{\rm{NUV}} - m_g = 3.5$ and another more extended and lower blue peak at around $m_{\rm{NUV}} - m_g = 2$. However, the red peak in Auriga is higher and more concentrated to larger values than ELVES, and the blue peak in ELVES is higher and less extended than that in Auriga. We note that we are using the FSPS default \texttt{MIST} isochrones, Chabrier IMF, and \texttt{MILES} spectral library, and that different isochrone models will produce different colours. 

Observationally, ELVES classify late- and early-type satellites by visually inspecting the morphology, which is not a direct method for classifying star-formation activity, unlike setting a threshold SFR, which is the criterion of separating star-forming and quenched satellites we used in Auriga. The discrepancy of the two colour distributions might be due to the different classification methods being used.

Finally, there might be sources other than the stars that emit NUV light, which means that the NUV-$g$ colour distribution does not cleanly distinguish the satellite star-formation activity.

Some observational studies also characterized quenching timescales, which are typically much harder to infer from observations. For example,
\citet{Fillingham2019} characterized the infall time for the population of MW satellite galaxies using \emph{Gaia} DR2 proper motion measurements from \citet{Fritz2018}. They found that the inferred quenching timescales for satellites of the MW within the mass range of $10^5 M_\odot \lesssim M_\star \lesssim 10^8 M_\odot$ is consistent with rapid cessation of star formation after infall \citep{Fillingham2015, Fillingham2016, Fillingham2018}, while satellites with mass $M_\star < 10^5 M_\odot$ are primarily quenched by reionization at early cosmic times, in agreement with our results.   Figures~\ref{fig:tau90_tinfall} and \ref{fig:tdelay_mass} show that ultra faint dwarfs (UFDs) at a critical mass scale of $M_\star \sim 10^5 M_\odot$ and have a quenching timescale $\tau_{90}$ close to the end of reionization, and that satellites of mass $10^5 M_\odot \lesssim M_\star \lesssim 10^7 M_\odot$ have either a negative or close to 0 delay time. 

Moreover, Figure~\ref{fig:CMD} shows that the Auriga simulations lack blue, low-luminosity satellites that exist in observations such as ELVES. To test if Auriga has these satellites at all, we computed NUV-$g$ colours for all satellites in the high-resolution region ($ < $1 Mpc from each host). However, we do not find any faint, blue satellites at all in the high-resolution region, suggesting that an issue of the model is at play here. In this faint regime of low mass satellites ($M_g > -12$), our model appears to fail to capture satellite star formation histories.  This appears not to be due to a lack of resolution, as L3 and L4 demonstrate the same trends, but due to the model itself.

One possible explanation for this discrepancy is the lack of low-temperature/molecular gas cooling in the stellar feedback and interstellar medium (ISM) model employed in the Auriga simulations. Auriga uses the subgrid ISM model of \citet{Springel.Hernquist2003}, which does not directly simulate the dense molecular gas but rather assumes it to be below the resolution of gas cells and treats it as in pressure equilibrium with the hot phase of the ISM. For faint satellites, when the UV radiation disturbs the gas, the effect of molecular gas self-shielding is likely underestimated. These cells therefore heat and stop forming stars more quickly than they should.  In low-mass satellites, this effect likely prevents extended, low-level star formation that would make the system bluer, causing the lack we see in Figure~\ref{fig:CMD}.  For example, in simulations of a dwarf galaxy heavily influenced by UV heating, \citet{Simpson2013} found that dense molecular gas could continue to self-shield, extending the SFH of an isolated dwarf system that would otherwise be quenched by external UV heating.
\subsection{Comparison with previous theoretical studies}
\label{sec:comp_models}
In this section, we compare our results of satellite infall time and colour distribution, satellite quenched fraction, and timescales to previous theoretical studies. In general, we found reasonably good agreement among different simulations with different underlying physics in terms of quenching timescales and quenched fraction across the mass range $10^5 M_{\odot} < M_\star < 10^{11} M_{\odot}$. We explore directly, for the first time, possible connections between satellite's position in the color--magnitude parameter space and quenching timescales and mechanisms. 

In recent years, theoretical studies have generally agreed upon the different quenching mechanisms operating on different mass scales. Lower-mass satellites ($M_\star < 10^6 M_\odot$) tend to quench as centrals \citep{Samuel2022, Fillingham2019,Simpson2018,Akins.etal.2021}, likely either by reionization at an early cosmic time, internal processes such as stellar feedback, or pre-infall environmental processes. By adding Auriga Level 3 data in our analysis, we probe into the regime of ultra faint dwarfs ($M_\star < 10^5 M_\odot$). In Figures~\ref{fig:tau90_tinfall} and \ref{fig:tdelay_mass}, we show that UFDs have a quenching timescale $\tau_{90}$ close to the end of reionization. For satellites within the mass range $10^5 M_\odot < M_\star < 10^6 M_\odot$, we show that they have a negative $t_{\rm{delay}}$ indicating that they are quenched before infall, in agreement with the scenario that low-mass satellites quench as centrals. 

Intermediate-mass satellites ($M_{\star} \sim 10^6 - 10^8 M_{\odot}$) tend to be quenched by environmental processes after infall, and the quenching timescale is rapid \citep{Wetzel2015b, Samuel2022, Akins.etal.2021}, i.e., $t_{\rm{delay}} \lesssim 2$ Gyr. A critical stellar mass scale of $10^8 M_{\odot}$ is identified by \citet{Akins.etal.2021}, where above this value satellites typically are resistant to quenching events and below this threshold satellites are quenched either by non-environmental processes such as reionization and stellar feedback or environmental processes such as ram pressure stripping.  Our results in Figure~\ref{fig:tdelay_mass} are consistent with this picture where we also see a transition of negative to positive $t_{\rm{delay}}$ around stellar mass $10^8 M_{\odot}$.

For massive satellites ($M_\star > 10^8 M_\odot$) the the quenching timescale is constrained by the gas depletion time ("starvation") and is not influenced by rapid environmental quenching effects \citep{Wetzel2015b}. Moreover, \citet{Joshi2021} found that in the case of satellites, dwarf systems with the highest satellite mass to host mass ratios have the most extended stellar mass assembly and the smallest $\tau_{90}$, which means that they can better resist environmental effects of their host and retain more gas in their reservoir, confirming our result in Figure~\ref{fig:tau90_tinfall} that satellites of higher stellar mass have a lower value of $\tau_{90}$. 

In this study, we focused on the intermediate luminosity range $-15 < M_g < -12$ where we see a transition of blue to red satellite galaxies in the CMD of both ELVES and Auriga satellites, and we identified environmental effects to be the dominant quenching mechanism for satellites in this luminosity range. In particular, we looked at the effect of ram pressure stripping in detail. Several other studies also looked at the role of ram pressure stripping on satellite quenching. \citet{Buck2019} followed \citet{Simpson2018} to calculate ram pressure and found that the sharp drop in gas fraction corresponds to the satellites approaching pericentre and thus experiencing an increased amount of ram pressure acting on their gas reservoir, which agrees with our results in Figure~\ref{fig:rp_illustration}. For lower mass satellite the ram pressure spikes up quickly after infall, shortly before their first pericentre and it quickly removes all the gas, whereas for higher mass satellites, they typically can resist the effect of ram pressure and still retain some gas after infall and the first pericentre passage \citep{Buck2019}, confirming our results for satellites of different mass in Figure~\ref{fig:rp_illustration}.

Several studies also identified ram pressure stripping, although not being the only quenching mechanism, is the dominate quenching mechanism for satellites with intermediate stellar mass $10^6 M_{\odot} \lesssim M_\star \lesssim 10^8 M_{\odot}$ \citep{Buck2019, Akins.etal.2021, Simpson2018}, which confirms our result in Figure~\ref{fig:trampeak} that lower-mass/faint systems ($M_g > -12$) typically reach a local ram pressure peak $< 1$ Gyr after infall, and their \tdelay\ is close to 0, which indicates that they are quenched mostly by ram pressure shortly after infall. Moreover, we find that intermediate luminosity/mass satellites that are quenched generally have a longer \tdelay. They are not quenched by ram pressure on the first pericentric passage, but might be quenched by later pericentric passages or starvation, which is indicated by the large scatter in Figure~\ref{fig:trampeak}.

Finally, striking commonalities in terms of the quenched fraction in different simulations are also reported by \citet{Sales.etal.2022}, indicating that this is a basic trend for every simulation no matter the underlying physics.

\section{Summary and conclusions}
\label{sec:summary}
This study is the first comparison between the new ELVES survey and the Auriga simulations. Our main results are summarized below:
\begin{itemize}
    \item[(i)] We confirm that simulations successfully capture intrinsic satellite galaxy properties such as the luminosity function (Figure~\ref{fig:LF}), quenched fraction across a wide range of satellite mass (Figure~\ref{fig:fquenched}), and the colour--magnitude distribution (Figure~\ref{fig:CMD}) from the ELVES and SAGA surveys. This demonstrates that we can trust simulations to probe into satellite observables such as colour and magnitude, and more importantly, we can use simulations to explore satellite properties such as infall time and different quenching timescales that are otherwise hard to obtain in observations. 
    \item[(ii)] We focus on the magnitude range $-15 < M_g < -12$ where ELVES found two distinct populations in terms of their NUV-$g$ colour. These two populations are also present in our colour--magnitude diagram (Figure~\ref{fig:CMD}) where there is a clear transition phase within this magnitude range. We also found two distinct populations in terms of their NUV-$g$ colour in the Auriga simulations (Figure~\ref{fig:colour_bimodal}), and we confirmed the results in \citet{Simpson2018} that there are also two distinct populations in terms of infall time in the Auriga simulations (Figure~\ref{fig:infall_bimodal})
    \item[(iii)] To better understand the origin of this transition phase in terms of satellite colour, we look into different satellite quenching timescales in the Auriga simulations. We found that low-luminosity satellites ($M_g > -12$) typically quench before infall (negative \tdelay, Figure~\ref{fig:tau90_tinfall}, Figure~\ref{fig:tdelay_mass}), likely by internal processes such as stellar feedback or by reionization. Luminous satellites ($M_g < -15$) are able to retain their gas reservoir even after infall into the host and thus are mostly still star-forming. Within the magnitude range $-15 < M_g < -12$, satellites are either star-forming or quenched, and for the quenched ones, they have a positive \tdelay which indicates that they are quenched by environmental processes after infall. Thus, we confirm that the two distinct populations in the NUV$-g$ colour distribution is caused by environmental quenching after infall. 
    \item[(iv)] We show that ram pressure stripping operates on a fast timescale ($\lesssim 1 $Gyr) for low-luminosity satellites ($M_g > -12$) upon infall in Figure~\ref{fig:trampeak}. For intermediate-luminosity satellites ($-15 < M_g < -12$), few of them are quenched when they reach a local ram pressure peak immediately after infall; they experience a  more prolonged quenching history either from later pericentre passages or starvation that gradually strips away all the gas. For luminous satellites, they are able to resist ram pressure and still retain their gas reservoir at the present day.  
    \item[(v)] Lastly, we found that the Auriga simulations lack a population of faint, blue satellites compared to observations such as ELVES. One possible explanation is the underestimation of  molecular gas self-shielding in low-mass systems due to the lack of molecular gas cooling in the Auriga ISM model. Future simulations will need to model the dense molecular gas more carefully to better reproduce observations in terms of the color--magnitude distribution. 
\end{itemize}

\section*{Acknowledgements}

The authors thank the anonymous referee for constructive suggestions on this manuscript. We acknowledge support from the University of Chicago's Research Computing Center and the Enrico Fermi Institute. The authors would like to thank Scott Carlsten, Jenny Greene, and the ELVES team for helpful discussion of the comparison between ELVES sample and the Auriga simulations. We also thank Phil Mansfield for comments on the figures, and Yao-Yuan Mao for kindly sharing the SAGA data. This work was supported by the Enrico Fermi Institute at the University of Chicago, by the National Science Foundation grants AST-1714658 and AST-1911111 and NASA ATP grant 80NSSC20K0512. FAG acknowledges support from ANID FONDECYT Regular 1211370 and by the ANID BASAL project FB210003. FAG acknowledges funding from the Max Planck Society through a “Partner Group” grant. Analyses in this paper were greatly
aided by the following free software packages: NumPy \citep{van.der.Walt2011}, SciPy \citep{jones_scipy:_2001}, Astropy \citep{astropy:2018}, and Matplotlib \citep{Hunter2007}. This research has also made extensive use of Astrophysics Data Service (\href{https://ui.adsabs.harvard.edu/classic-form/}{\tt{ADS}}) and \href{https://arxiv.org/}{\tt{arXiv}} preprint repository

\section*{Data Availability}

The data underlying this article will be shared on reasonable request to the authors.

\bibliographystyle{mnras}
\bibliography{references} 

\bsp
\label{lastpage}
\end{document}